\documentstyle[12pt,openbib]{article}
\def\J{{\cal J}}
\def\h{h_{ab}}
\def\p{\pi^{ab}}

\def\P{\pi_{ab}}

\def\conf{ (\h,\p,\phi_i,P_i )}

\def\J{\cal{J}}

\begin{document}

\title{ Quasi-Asimptotically Flat Spacetimes and Their ADM Mass}
 
\author{{\normalsize  Ulises Nucamendi\thanks{email: ulises@fis.cinvestav.mx}
and Daniel Sudarsky\thanks{email: sudarsky@xochitl.nuclecu.unam.mx}} \\
{\normalsize $^\dagger$ Instituto de Ciencias Nucleares}\\
{\normalsize Universidad Nacional Aut\'onoma de M\'exico}\\
{\normalsize A. P. 70-543, M\'exico, D. F. 04510,  M\'exico.} \\
{\normalsize $^*$ Departamento de F\'{\i}sica}\\
{\normalsize Centro de Investigaci\'on y de Estudios Avanzados del I.P.N.}\\
{\normalsize A. P. 14-741, M\'exico, D. F. 07000,  M\'exico.} \\
}
\date{}
\maketitle
\begin{abstract}

We define spacetimes that are asymptotically flat,
except for a deficit solid angle $\alpha$,
and present a definition of their
``ADM" mass, which is finite
for this class of spacetimes, and, in particular,
coincides with the value of
the parameter $M$ of the global monopole spacetime
studied by Vilenkin and Barriola 
\cite{Vilenkin}. Moreover, we show that the
definition is coordinate
independent, and explain why it can,
in some cases, be negative.
\end{abstract}

\medskip
\section{Introduction}
\setcounter{equation}{0}
\medskip
Spacetimes that are asymptotically flat (A.F.),
and the properties of their associated
ADM mass (or, more generally, four momentum)
have been studied
exhaustively \cite{ADM}, \cite{Geroch1},
\cite{Geroch2},
\cite{Ashtekar-Hansen}, \cite{Ashtekar},
as these are taken as the natural idealizations of those
spacetimes that represent isolated objects
in General Relativity. However, these are not
the only  idealizations that are of interest,
since, in fact, our own universe, having a
nonzero average density, is not asymptotically
flat, and, moreover, localized solutions that
can not be naturally accommodated
within the asymptotically flat framework,
can naturally be
considered as describing regions of our universe.
Such is the case for, the so called
global monopoles studied by
Vilenkin and Barriola \cite{Vilenkin}, for which
the energy density drops off only as $1/r^2$.
Thus, these monopole spacetimes are not asymptotically flat in the standard sense.
In particular they have a divergent value for  
the expression  that would have defined their 
ADM mass. However, it is clear that for
$r$ sufficiently large, that the
monopole density
becomes smaller than the mean matter density of
the universe, the particular form of its
subsequent rate of decay is of no consequence
whatsoever. Thus,
it should make physical sense to seek a notion of the asymptotic
behavior of a spacetime that is appropriate for the description of
such type of
``quasi-isolated objects" and to study
the properties that can be defined for them.

In this paper, we exhibit the first steps of
such a program for the class
of spacetimes that are asymptotically flat,
but for a deficit solid angle $\alpha$
(A.F.D.A.$\alpha$),
which we define more specifically below.
This class includes the global monopole
 of Vilenkin and Barriola \cite{Vilenkin} and
some  perturbations thereof.
We give a definition of mass that is
finite for the class of (A.F.D.A.$\alpha$)
spacetimes, and, moreover, coincides with
the value of the parameter $M$ of the
global monopole solution. Finally,
we briefly
discuss some aspects of the asymptotic
symmetry group of these spacetimes.

Previous works
along these general lines include the
analysis of Abbott and Deser \cite{Abott}
of the canonical mass for asymptotically
De Sitter and Anti De Sitter spacetimes
(See also \cite{Nester}). In that study,
the authors encountered problems related
to the fact that in the De Sitter case the
``mass" can only be associated with 
an horizon  sized region of a Cauchy
hypersurface, and in the Anti De Sitter
case there is no Cauchy hypersurface at all.
In the present work, we do not encounter
those problems.

We shall adhere to the following conventions
on index notation in this paper: Greek indices
($\alpha$, $\beta$, $\mu$, $\nu$,...) range
from $0$ to $3$, and denote tensors on
(four-dimensional) spacetime. Latin indices,
alphabetically located after the letter
i (i,j,k,...), denote internal indices in the
space of scalar fields, and range from $1$ to $3$;
whereas Latin indices from the beginning of
the alphabet (a,b,c,d,...) range from $1$ to $3$,
and denote tensors on a spatial
hypersurface  $\Sigma$.

Metric tensors are employed throughout
the paper: {\bf g} denotes the spacetime
 metric, {\bf h} denotes the metric on a
spatial hypersurface $\Sigma$.
The corresponding covariant derivatives are
denoted $\nabla$, for the metric {\bf g}, and $D$
for the metric {\bf h}.

The signature of the metric {\bf g} is $ (-,+,+,+)$.
Geometrized units, for which
$G_N=c=1$ are used in this paper.

\medskip
\section{The Global Monopole Spacetime}
\medskip

The theory of a scalar field with spontaneously broken
internal $O(3)$
symmetry, minimally
coupled to gravitation, is described
by the action:
\\
\begin{equation}
S = \int \sqrt{(-g)}
[(1/16\pi) R - (1/2)\nabla^{\mu}\phi_i
\nabla_{\mu} \phi_i -V(\phi)]  d^{4} x.
\label{(1.1)}
\end{equation}
\\
where R is the scalar curvature of the
spacetime metric, $\phi_i$ is a
triplet of scalar fields, and
$V(\phi)$, is potential depending only
on the magnitude $\phi =(\sum _i \phi_i^2)^{1/2}$,
which we will usually take to be the ``Mexican Hat" 
$V(\phi) = (\lambda/4)
(\phi^2 - v^2)^2$.

We are interested in spacetimes with topology
$\Sigma \times R$, where
$\Sigma $ has the topology of $(R^3 - B) Ê\cup C$,
with $B $ a 3-ball, and $C$ a compact manifold with
$S^2$ boundary.
The requirement that $ \phi \to v$ in the
asymptotic regions
separates the configuration space into
topological sectors according to the
winding number
of the asymptotic behavior  of $\phi_i$.
We will focus on the sector with winding
number one, corresponding to the
asymptotic behavior characteristic of the
Hedgehog ansatz:

\begin{equation}
\phi_i
\approx  v  x_i /r.
\label{(1.2)}
\end{equation}
\\
where the $x^i$  are  asymptotic cartesian
coordinates.
Within this sector, there is a static,
spherically symmetric solution \cite{Vilenkin}
with metric
given by:
\\
\begin{equation}
 ds^2 = - B(r) dt^2 + A(r) dr^2
 + r^2  ( d\theta^2
+ \sin^2 (\theta) d\varphi^2 ),
\label{solution}
\end{equation}
\\
and scalar field
\begin{equation}
\phi_i = v f(r)  x_i /r
\label{escalar}
\end{equation}
\\
and with the following asymptotic behavior of,
\\
\begin{equation}
B=A^{-1} = 1 - \alpha - 2 M/r + O(1/r^2),
\quad  f \approx 1 + O(1/r^2)
\label{CA}
\end{equation}
\\
where $\alpha= 8\pi v^2$.
\\
Redefining the $r$ and $t$ coordinates
as $r \to (1-\alpha)^{1/2} r$ and
$t \to (1-\alpha)^{-1/2} t$, respectively,
and defining $\widetilde M = M (1-\alpha)^{-3/2}$,
we obtain the asymptotic form for the metric:
\\
\begin{equation}
 ds^2 = - (1-2\widetilde M /r) dt^2 +
(1-2\widetilde M /r)^{-1} dr^2
 +(1-\alpha) r^2  ( d\theta^2
+ \sin^2 (\theta) d\varphi^2 ).
\end{equation}
\\
The parameter $\widetilde M$ has previously
been associated with the mass of the
configuration ( despite the fact that the
ADM mass formally diverges)
because it can be seen that the proper
acceleration of the  ($\theta,
\varphi, r$) = constant
world lines is $a= - \widetilde M/(r(r-2\widetilde M))$.
Thus
$\widetilde M$ plays a role of a Newtonian mass.
However, let us emphasize that the ADM mass
of the solution is not defined and that the
formal application of the ADM formula actually
diverges. It is also rather unexpected 
that in the specific solution $\widetilde M$ turns
out to be negative \cite{Vilenkin2}.

Another intriguing question is posed by
the fact that in \cite{Sork} and \cite{us1}Ê a
very close
connection between staticity and
extrema of mass was found, and thus it
is somehow
surprising that here we find static solutions
that do not seem to be extrema of anything.

We will see that these points seem to be
completely resolved by the introduction of
the notion of
 A.F.D.A $\alpha$ spacetime and the new
definition of ADM mass that is
appropriate for them.

\medskip
\section{The New Class of Spacetimes and Their ADM Mass}
\medskip

We start with the definition of a standard
spacetime which is going to play the
role that Minkowski spacetime plays for the
case of asymptotically flat spacetimes;
namely, it is going to be used in the specification
of the asymptotic behavior defining the
the new class of spacetimes, and as a benchmark
for the definition of the ``ADM mass".
\\
We take the spacetime  to be
$(R^3-\vec 0) \times R$ ( $\vec 0$ is the
origin of $R^3$) with the metric :
\\
\begin{equation}
 ds^2 = g_{\mu\nu}^0 dx^\mu dx^\nu =
-dt^2 + dr^2 + r^2 (1-\alpha) ( d\theta^2
+ \sin^2 (\theta) d\varphi^2 ).
\label{standard}
\end{equation}
\\
We will call this spacetime the standard
asymptotically-flat-but-for-a-deficit-angle
$\alpha$ spacetime or (S.A.F.D.A $\alpha$).

It can be considered as the global monopole
solution in the limit when
$\lambda \to \infty$,
or, more precisely, as the solution for
the model in which the potential has
been replaced by the constraint
$\sum_{i=1}^{3} \phi_{i} \phi_{i} = v^2$.

The asymptotic features of this spacetime
can be analyzed
by carrying out a compactification analogous
to the standard
 conformal  compactification of Minkowski
spacetime \cite{Geroch1},\cite{Ashtekar-Hansen},
\cite{Ashtekar},
\cite{Penrose1},\cite{Penrose2}.
In fact,
we can introduce new coordinates $u$,$v$
according to,
\\
\begin{equation}
u = t + r,  \quad \quad v = t - r.
\end{equation}
\\
In these coordinates, the S.A.F.D.A $\alpha$
 metric takes the form,
\\
\begin{equation}
 ds^2 = -dudv +
\frac{1}{4} (v-u)^{2} (1-\alpha)
( d\theta^2
+ \sin^2 (\theta) d\varphi^2 ),
\label{cambio}
\end{equation}
\\
which is seen to be conformally related
to the metric
\\
\begin{equation}
\tilde{ds}^2  = -dT^{2} + dR^2 +
(1-\alpha) (\sin^2 R)
 ( d\theta^2 + \sin^2 (\theta) d\varphi^2 ),
\label{conformal}
\end{equation}
\\
i.e., $\tilde{ds}^2 = \Omega^2 ds^2$
with conformal factor
$ \Omega^2 = \frac{4}{(1 + v^2)(1 + u^2)}$,
 where
\\
\begin{equation}
T = \arctan v + \arctan u, \quad R =
\arctan v - \arctan u,
\end{equation}
\\
and where $T,R$ have the following ranges,
\\
\begin{equation}
-\pi < T + R < \pi,
\quad  -\pi < T - R < \pi,
\quad 0 < R < \pi.
\end{equation}
\\
This spacetime can be extended to
$T = - \pi + R$ and $T = \pi - R$ for
 $R \in (0,\pi)$, which correspond to
past null infinity $\J^{-}$, and
future null infinity $\J^{+}$, respectively.
Unfortunately, this spacetime can not be
extended
to ($T = 0$, $R = \pi$), which would have
corresponded to
spatial infinity $\imath^{0}$, because
here there is a true singularity that
is evidenced
by the fact that the scalar curvature
diverges at $R = \pi$
(in fact, the curvature scalar is
 $ (6-4\alpha+2\alpha\cot^2R)/(1-\alpha)$).
There is a similar
singularity at $r = 0$ present
in the real spacetime, but we will not be
concerned with it because we will be
interested in spacetimes which are similar
to the S.A.F.D.A $\alpha$ spacetime only in
the asymptotic region, and these
will include many regular spacetimes in 
which this conical singularity will
be ``smoothed out".
In the S.A.F.D.A $\alpha$ spacetime,
the presence of the singularity is
the price that we pay in order to have
a very simple spacetime to take as
the standard one. We could have equally
chosen any of the ``smoothed out"
spacetimes as the standard, but
there seems to be no canonical choice.

The fact that we can introduce the concepts
of $\J^{-}$ and $\J^{+}$, but not
$\imath^{0}$, suggests that in the terminology of \cite{Hawk1}
these spacetimes
would be ``asymptotically
simple", but not ``asymptotically empty". \\
\\
{\bf Definition}: We will say that a spacetime
(or a spacetime region) $(M, g_{\mu\nu})$,
with topology $(R^3 -B)\times R$, is
asymptotically
flat, but for a deficit
angle $\alpha$  (A.F.D.A $\alpha$), if
there exist coordinates
 $( t, r,\theta, \varphi)$, for which
the metric can be written as
\begin{equation}
g_{\mu\nu}= g_{\mu\nu}^0 + \tilde g_{\mu\nu},
\label{defmet1}
\end{equation}
\\
where $\tilde g_{\mu\nu}$ has the form:
\\
\begin{eqnarray}
\tilde g_{\mu\nu} dx^\mu dx^\nu & = &
a_{tt} dt^2 + a_{rr}
dr^2 + 2a_{rt} drdt \nonumber \\
&& + \, r^2
[a_{\theta\theta} d\theta^2 + a_{\varphi\varphi}
\sin^2 (\theta) d\varphi^2
+ 2a_{\theta\varphi} \sin (\theta)
d\theta d\varphi] \\
&& + \, 2r [a_{t \theta} dtd\theta
+ a_{r \theta} drd\theta] +
 2r [a_{t \varphi} \sin (\theta) dtd\varphi +
 a_{r \varphi} \sin (\theta) drd\varphi],
\nonumber
\label{defmet2}
\end{eqnarray}
\\
with the functions $a_{\mu\nu} \approx O(1/r)$
 (note that the $a_{\mu\nu} $ depend on the choice of background). \\
\\
We review now the ``3 + 1"  hamiltonian
formulation of the
Einstein-Scalar (E-S)
theory, analogous to that given in \cite{us1},
 and proceed to specialize
considerations to the phase space of regular,
 asymptotically-flat-but-for-a-deficit-angle
 $\alpha$ initial data.
\\
Initial data in  E-S theory consists of the
specification of the fields $\conf $ on a
three-dimensional manifold, $\Sigma$.
Here $\h$ is a Riemannian metric on $\Sigma$,
$\phi_i $  
is the $i^{th}$ scalar field component ,
$\p $ is the canonically conjugate momentum
to $\h$, and $P_i$ is the
momentum canonically conjugate to $\phi_i$.
\\
Einstein-Higgs theory is a theory with
constraints. On a hypersurface, $\Sigma$, 
the allowed initial data are restricted to
those that at each point $x \in \Sigma $ satisfy

\begin{eqnarray}
0 & = & {\cal C}_0=\sqrt h
[-^{(3)}R + 1/h(\P\p -(1/2) \pi^2)]
\nonumber \\
&& \quad \quad + (P_iP_i/2\sqrt h) + \sqrt h
 (  (1/2) h^{ab} D_a\phi_i D_b\phi_i + V(\phi)),  \\
&& \nonumber \\
0 & = & {\cal C}_a=
-2\sqrt h D_b (\pi_a^{\ b} /\sqrt h)
+P_i D_a \phi_i.
\label{(2.13)}
\end{eqnarray}
\\
A fixed volume element $\eta_{abc}$  for
the manifold $\Sigma$ is assumed to be given,
and $h$ relates the volume element
$\epsilon_{abc}$ corresponding to the metric
$h_{ab}$ to the former through
$\epsilon_{abc}=\sqrt h \eta_{abc}$.
\\
The equations of motion of E-S theory can be
derived from a Hamiltonian $H$.
(See also \cite{Tietl}).
\\
\begin{equation}
H = \int_{\Sigma} (N^\mu {\cal C}_\mu).
\label{(2.14)}
\end{equation}
\\
where the fixed volume element $\eta_{abc}$ is
understood in all volume integrals over $\Sigma$.
\\
$N^\mu =(N^0,N^a)$ corresponds to
the ``lapse"  function and ``shift'' vector,
respectively, of the
foliation of the ``evolved spacetime".
Recall that $N^\mu$ are not dynamical
variables, and can be chosen arbitrarily .

A general variation of the initial data will
produce a variation
in the Hamiltonian that can be written as:
\\
\begin{equation}
\delta H = \int_{\Sigma} (
{\cal P}^{ab}\delta \h +
{\cal Q}_{ab}\delta \p +
{\cal R}^{i} \delta \phi_i+
{\cal S}^i \delta P_i) + Surface \quad terms.
\label{HalVar}
\end{equation}
\\
The evolution equations can be obtained
from Hamilton's principle, if we
restrict consideration to variations of
compact support,
\\
\begin{equation}
\dot\p  =  -{\cal P}^{ab},\quad
\dot\h  =  {\cal Q}_{ab}, \quad
\dot P_i  =  -{\cal R}_{i},\quad
\dot \phi_i  =  {\cal S}_i, 
\end{equation}
\\
where specific expressions for
${\cal P}^{ab},{\cal Q}_{ab},{\cal R}_i$,
and ${\cal S}_i$, and the surface terms
are given by:

\begin{eqnarray}
{\cal P}^{ab} & = & \sqrt h N^0
[^{(3)}R^{ab} - (1/2) ^{(3)}R h^{ab}]
- (N^0/2\sqrt h) h^{ab}
[\pi_{cd}\pi^{cd}  -(1/2) \pi^2]  \nonumber \\
&& + 2 (N^0/\sqrt h)
[\pi^{ac}\pi_{c}^{b} - (1/2) \pi\pi^{ab}] -
\sqrt h [D^{a}D^{b}N^0 - h^{ab} D^{c}D_{c}N^0]
\nonumber \\
&& - \sqrt h D_{c}(N^{c} \pi^{ab}/\sqrt h)
+ 2 \pi^{c(a}D_{c}N^{b)} -
 N^0 h^{ab} (P_iP_i/2\sqrt h) \nonumber \\
&& + (1/2) \sqrt h N^0 h^{ab}
[(1/2) h^{cd} D_{c}\phi_i D_{d}\phi_i +
V(\phi)] - \sqrt h N^0 D^{a}\phi_i D^{b}\phi_i ,
\\
&& \nonumber \\
&& \nonumber \\
{\cal Q}_{ab} & = & (2N^0/\sqrt h)
[\pi_{ab} - (1/2)h_{ab} \pi] + 2D_{(a}N_{b)}, 
\\
&& \nonumber \\
&& \nonumber \\
{\cal R}_i & = & \sqrt h N^0
[ \frac{\partial V(\phi)}{\partial \phi_i} -
 D_{a}D^{a}\phi_i ]
- \sqrt h D_{b}(N^0)D^{b}(\phi_i)
- \sqrt h D_{a}
 ( N^{a} P_{i}/\sqrt h ), \\
&& \nonumber \\
&& \nonumber \\
{\cal S}_i & = &
N^{a} D_{a}\phi_i + N^{0} P_{i}/\sqrt h 
\label{ecsmov},
\end{eqnarray}

\begin{eqnarray}
surfaces \quad terms & = &
\int_{\partial\Sigma} dS_{a}
[ (D_{b}N^0) (h^{ac}h^{bd} - h^{ab}h^{cd})
\delta h_{cd} \nonumber \\
&& - N^0 (h^{ac}h^{bd} - h^{ab}h^{cd})
D_{b}(\delta h_{cd}) -
 2 (N^b/\sqrt h) \delta \pi^{a}_{b}
\nonumber \\
&& + (N^a/\sqrt h) \pi^{bc} \delta h_{bc}
+ N^0 (D^{a}\phi_i)
\delta\phi_i
+ (N^a/\sqrt h) P_i \delta\phi_i ].
\label{surface}
\end{eqnarray}
\\
These eqs. are known to be equivalent to
the four-dimensional E-S 
equations. However, as pointed out by
Teitelboim\cite{Tietl}, this is not a
satisfactory application of Hamilton`s
principle,
which must consider unrestricted
variations within the phase space.
If we specify the phase space to be
that of asymptotically flat regular
initial data, and consider evolution that
corresponds asymptotically to a
``time translation", the problem is resolved
 by adding a surface term to the
Hamiltonian. The surface term is just
the ADM mass.
If we now  specify phase space to be
that of (A.F.D.A.$\alpha$) regular initial
 data, which we will define below, it will
turn out that, again, a surface
term can be added to the Hamiltonian that
results in a satisfactory Hamiltonian
for the application of Hamilton`s principle,
for evolution corresponding
asymptotically to a ``time translation", this is:

\begin{eqnarray}
N^0 & \approx & 1 + O(1/r), \\
&& \nonumber \\
N^a & \approx & O(1/r).
\end{eqnarray}
\\
{\bf Definition}: We will call A.F.D.A.$\alpha$
regular initial data
 to a specification of the fields $\conf $ on a
three-dimensional
manifold, $\Sigma$, which fulfills the following
conditions at infinity,

\begin{eqnarray}
h_{ab} & \approx & h_{ab}^0 + \delta h_{ab},
\quad \quad  \delta h_{ab}
\approx O(1/r),  \\
&& \nonumber \\
h_{ab,c} & \approx & O(1/r^2), \\
&& \nonumber \\
\pi^{ab} & \approx & O(1/r^2),  \\
&& \nonumber \\
\phi_i & \approx & v x_i /r + O(1/r^2),  \\
&& \nonumber \\
P_i & \approx & O(1/r),
\end{eqnarray}
\\
where $h_{ac}^{0}$ is the S.A.F.D.A
$\alpha$ spatial metric defined
for the hypersurface $t=$ constant
(\ref{standard}), and $D_{b}^{0}$ the
covariant derivative associated with it.

The surface term, whose variation will
cancel the nonvanishing
surface term in (\ref{surface}), when
asymptotic conditions are imposed, is:
\\
\begin{equation}
16 \pi (1-\alpha) M_{ADM\alpha} =
\int_{\partial\Sigma}
dS_{a} ( h^{0ac} h^{0bd} - h^{0ab} h^{0cd} )
 D_{b}^0 (h_{cd}).
\label{masa}
\end{equation}
\\
This is clearly the natural generalization of
the ADM mass, (in fact, it looks just like the
usual ADM formula, but with the quantities
associated with the flat metric replaced by the
S.A.F.D.A.$\alpha$ metric), and just like this,
it is the numerical value of the true Hamiltonian
( a true generator of ``time translations");
so, it is natural to interpret this as the mass
(or energy) of the A.F.D.A.$\alpha$ spacetimes.

This interpretation is reinforced by the
fact that, when applied to the Global
Monopole solution, it
yields the value:

\begin{equation}
M_{ADM \alpha} = \tilde M.
\end{equation}

\medskip
\section{The new mass formula is well defined}
\medskip
The problem, in principle, with the above
definition stems from the fact
that the formula (\ref{masa}) involves
geometries in two different spaces.
Specifically, we are using covariant
derivatives associated with one metric,
and applying it to a second one. Thus,
what we have in fact
is the following:
A standard-setting Riemmanian Manifold
$(\Sigma^0, {\bf h}^0)$, a test
Riemmanian Manifold
$(\Sigma, {\bf h})$,  and a mapping
(we assume it is a diffeomorphism)
 $\Phi:\Sigma^0\to \Sigma $. The metric
appearing in eq. ($\ref{masa}$)
is actually $\Phi^* ({\bf h})$. So, it is
not clear, in principle, that the
definition of
$M_{ADM\alpha}$ does not depend on $\Phi$.
Therefore, we need to consider two
such diffeomorphisms
$\Phi_1$, $\Phi_2:\Sigma^0 \to \Sigma$
(which must preserve the asymptotic form
of the metric
${\bf h}$, written in the coordinates
$(r,\theta, \varphi)$, associated with
the S.A.F.D.A $\alpha$), and see that the value
of ADM$\alpha$ mass is the same for
$\Phi_1$ and $\Phi_2$.
This amounts to considering a change
of variables ($x^{a} \to x'^{a} = y^{a}$),
which preserves the asymptotic form of
${\bf h}$, and then dropping the primes
and substituting directly into the
expression (\ref{masa})
(without making any change in the variables
there, i.e., without change
$h_{ab}^{0}$ and $D_{b}^{0}$).
It will be convenient to write everything
in a chart $\psi:\Sigma^0 \to R^3$ on
 $\Sigma^0$ with $x^{a} =\psi^a(p)$
Cartesian coordinates defined
 by means of their usual relation with
spherical coordinates $(r,\theta, \varphi)$,
$p \in \Sigma^0$. In these coordinates,
the S.A.F.D.A $\alpha$ metric
${\bf h}^0$ is written as,
\\
\begin{equation}
h^{0}_{ab} =
[(1-\alpha) \delta_{ab} + \alpha x^{a}x^{b}/r^2].
\label{met} 
\end{equation}
\\
We will use the following relation in the paper,
\\
\begin{equation}
x_{a} = h^{0}_{ab} x^{b} = x^{a},
\end{equation}
\\
then we can write,
\\
\begin{equation}
h^{0}_{ab} =
[(1-\alpha) \delta_{ab} + \alpha x_{a}x_{b}/r^2].
\label{metr} 
\end{equation}
\\
Using the charts
$\psi \circ \Phi_1^{-1}:\Sigma \to R^3$
and $\psi \circ \Phi_2^{-1}:\Sigma \to R^3$
(strictly speaking, these maps will be defined
only in the asymptotic region),  we can write
 $\Phi^{*}_{1}({\bf h})$, and
$\Phi^{*}_{2}({\bf h})$ in the ''cartesian"
coordinates $x^{a}=\psi^{a} \circ \Phi_1^{-1}(q)$,
and $y^{a}=\psi^{a} \circ \Phi_2^{-1}(q)$,
for $q \in \Sigma$, which we will refer to
 as $h_{(1)ab}$ and $h_{(2)ab}$, respectively,
thus we have:

\begin{equation}
h_{(1)ab} =
[(1-\alpha) \delta_{ab} +
\alpha x_a x_b/r^2 + A_{ab}],
\label{met1}
\end{equation}

\begin{equation}
h_{(2)ab} =
[(1-\alpha) \delta_{ab} +
\alpha y_a y_b/r'^2 + B_{ab}],
\label{met2}
\end{equation}
\\
here we have,
\\
\begin{equation}
A_{ab} \approx O(1/r),
\quad \, B_{ab} \approx O(1/r'), \quad \,
\frac {\partial A_{ab}} {\partial x^{c}}
\approx O(1/r^2), \quad \,
\frac {\partial B_{ab}} {\partial y^{c}}
\approx O(1/r'^2),
\label{condi}
\end{equation}
\\
where $r^2 = \sum_{a=1}^{3} x^{a}x^{a}$,
$r'^2 = \sum_{a=1}^{3} y^{a}y^{a}$.
The coordinates ${x^{a}}$ and ${y^{b}}$ are
related by the diffeomorphism
$\{x^{a}\} = \psi^{a} \circ \Phi_1^{-1}
\circ \Phi_2 \circ \psi^{-1} \{y^{b}\}$.
\\
The proof that the value of ADM$\alpha$
mass is independent of
diffeomorphisms $\Phi$ that preserve
the asymptotic form of the metric
${\bf h}$ is basically a repetition
(with some modifications) of the
proof of the analogous
statement for the ADM mass of
asymptotically
flat spacetimes which
has been given in \cite{piotr}.  \\
\\
{\bf Lemma 1}: The conditions (\ref{condi})
imply that $h_{(1)ab}$ and
$h_{(2)ab}$ are uniformly elliptic, this is,
for $r,r'$ sufficiently large,
there exist positive constants $C_{1},C_{2}$,
such that, $\forall$ vector $ \, X^{a}$,
\\
\begin{equation}
C_{i}^{-1} \sum_{a=1}^{3}
X^{a}X^{a} \leq h_{(i)ab} X^{a}X^{b}
\leq C_{i} \sum_{a=1}^{3} X^{a}X^{a},
\label{eliptica}
\end{equation}
\\
where $i = 1,2$. \\
\\
Proof: We will prove it, say,
for $h_{(1)ab}$.
First we note that,
\\
\begin{equation}
\exists C>0, \exists r_{0}>1
\quad \mbox{so \,\, that}
\quad \forall r>r_{0},
\quad |A_{ab}|\leq \frac {C}{r}.
\label{des2}
\end{equation}
\\
From Schwartz's inequality, we obtain,
\\
\begin{equation}
\sum_{a,b=1}^{3}
\frac{x_{a}X^{a}x_{b}X^{b}}{r^2} \leq
(\sum_{c=1}^{3} X^{c}X^{c})^{1/2}
(\sum_{b=1}^{3} X^{b}X^{b})^{1/2}
= \sum_{c=1}^{3}(X^{c}X^{c}).
\end{equation}
\\
From the above inequalities, using
$(1-\alpha)>0$, $\alpha>0$,
and choosing $D>(1-\alpha)$, we conclude,

\begin{equation}
\exists S_{1} = (D + \alpha + 9C)>0
\quad \mbox{so \,\, that}
 \quad \forall r>r_{0}>1,
\quad h_{(1)ab} X^{a}X^{b} \leq S_{1}
\sum_{a=1}^{3} X^{a}X^{a}.
\label{hiper1}
\end{equation}
\\
From inequality (\ref{des2}), we obtain,
\\
\begin{equation}
\forall r>r_{1}>r_{0}, \quad
- \frac {9C}{r_{1}}
\sum_{c=1}^{3} X^{c}X^{c}
 \leq -\frac {9C}{r}
\sum_{c=1}^{3} X^{c}X^{c} \leq
\sum_{a,b=1}^{3} X^{a}X^{b}A_{ab}.
\end{equation}
\\
Using the above inequalities, and
choosing $r_{1} \geq 9C/(1-\alpha)$, and $r_{1}>r_{0}$, we have,
\\
\begin{equation}
\exists S_{2} = (1-\alpha
- \frac {9C}{r_{1}})>0, \quad
\mbox{so \,\, that}
\quad \forall r>r_{1}, \quad S_{2}
\sum_{a=1}^{3} X^{a}X^{a}
\leq h_{(i)ab} X^{a}X^{b}.
\label{hiper2}
\end{equation}
\\
We take $r_{2} = max(r_{0},r_{1})$,
$S = max(S_{1},S_{2})$, and
then eqs. (\ref{hiper1}), (\ref{hiper2})
prove the lemma. \\
\\
{\bf Lemma 2}: Let $\{x^{a}\}$ and
$\{y^{a}\}$ be coordinate systems on
$\Sigma$ for which the
metrics $\Phi^{*}_{1}({\bf h})$,
$\Phi^{*}_{2}({\bf h})$, respectively,
preserve the asymptotic
form (eqs. (\ref{met1}), (\ref{met2})),
and such that the diffeomorphism
$\{x^{a}\} = \psi^{a} \circ \Phi_1^{-1}
\circ \Phi_2 \circ \psi^{-1} \{y^{b}\}$
is at least twice differentiable.
Then, this diffeomorphism, and its inverse
have the form,

\begin{equation}
x_{a}(y) = x^{a}(y) =
\sum_{b=1}^{3} W^{a}_{b} y^{b} + \eta^{a}(y),
\label{dif1}
\end{equation}

\begin{equation}
y_{a}(x) = y^{a}(x) =
\sum_{b=1}^{3} (W^{-1})^{a}_{b} x^{b}
+ \zeta^{a}(x),
\label{dif2}
\end{equation}
\\
where,
\\
\begin{equation}
\eta^{a} \approx O(1),
\quad \, \zeta^{a}\approx O(1), \quad \,
\frac {\partial \eta^{a}(y)} {\partial y^{b}}
\approx O(1/r), \quad \,
\frac {\partial \zeta^{a}(x)} {\partial x^{b}}
\approx O(1/r), \quad
W^{a}_{b} \in SO(3).
\end{equation}
\\
\\
Proof: We can regard the diffeomorphism
as defining a change of coordinates,

\begin{equation}
h_{(2)ab}(y) = h_{(1)cd}(x(y))
\frac {\partial x^{c}} {\partial y^{a}}
\frac {\partial x^{d}} {\partial y^{b}},
\label{Trans1}
\end{equation}

\begin{equation}
h_{(1)ab}(y) = h_{(2)cd}(y(x))
\frac {\partial y^{c}} {\partial x^{a}}
\frac {\partial y^{d}} {\partial x^{b}}.
\label{Trans2}
\end{equation}
\\
Introducing $\delta_{ab} = e_{a(k)}e_{b(k)}$
(where $e_{a(k)}$ is
the $a^{th}$-component of the $k^{th}$
unitary cartesian vector),
contracting (\ref{Trans1}), (\ref{Trans2})
with $\delta_{ab}$ and using the
property (\ref{eliptica}), we obtain,

\begin{eqnarray}
\frac{1}{C} \sum_{a,c=1}^{3}
\frac {\partial x^{c}} {\partial y^{a}}
\frac {\partial x^{c}} {\partial y^{a}}
\leq \sum_{a=1}^{3} h_{(1)cd}(x(y))
\frac {\partial x^{c}} {\partial y^{a}}
\frac {\partial x^{d}} {\partial y^{a}}
& = & \delta_{ab} h_{(2)ab}(y) =
h_{(2)ab} e_{a(k)}e_{b(k)} 
\nonumber \\
& \leq & C \sum_{c=1}^{3}
e_{c(k)}e_{c(k)} = 3C,
\end{eqnarray}
\\
and an analogous expression with
$(2) \leftrightarrow (1)$.
This shows that the derivatives of
$x(y)$, $y(x)$ are uniformly bounded,
\\
\begin{equation}
\left|
\frac {\partial x^{a}} {\partial y^{b}} 
\right|
\leq \sum_{a,c=1}^{3}
\frac {\partial x^{c}} {\partial y^{a}}
\frac {\partial x^{c}} {\partial y^{a}}
\leq 3C^2, \quad \,
\left|
\frac {\partial y^{a}} {\partial x^{b}} 
\right|
\leq \sum_{a,c=1}^{3}
\frac {\partial y^{c}} {\partial x^{a}}
\frac {\partial y^{c}} {\partial x^{a}}
\leq  3C^2.
\label{acotado}
\end{equation}
\\
Thus eq. (\ref{acotado}) implies, for
$r,r'$ sufficiently large, and a curve
$\Gamma$ on
$\Sigma$($\Gamma:[0,1] \mapsto \Sigma$,
$\Gamma(0) = r_{0}$,
$\Gamma(1) = r$, with
$\Gamma(t) \geq r_{0}$),
that,
\\
\begin{equation}
\left| y^{a}(x) \right| = \left|
\sum_{b=1}^{3} \int_{\Gamma}
\frac {\partial y^{a}}{\partial x^{b}}
dx^{b} \right| \leq 3C r(x),
\end{equation}
\\
\begin{equation}
\left| x^{a}(y) \right| = \left|
\sum_{b=1}^{3} \int_{\Gamma}
\frac {\partial x^{a}}{\partial y^{b}}
dy^{b} \right| \leq 3C r'(y).
\end{equation}
\\
Using the above, we can conclude that
there is a constant $C'$, such that,
\\
\begin{equation}
r'(y(x)) \leq C' r(x),
\quad \quad r(x(y)) \leq C' r'(y),
\label{des3}
\end{equation}

or, in other words,

\begin{equation}
r'(y(x))/C' \leq r(x) \leq C' r'(y(x)),
\quad  r(x(y))/C' \leq r'(y) \leq C' r(x(y)).
\label{desigualdad}
\end{equation}
\\
Then, any function $f(x(y)) \approx O(1/r)$
if and only if $f(x(y)) \approx O(1/r')$.
Also from eqs. (\ref{condi}), (\ref{acotado}),
we conclude that the functions
$A_{ab}$,$B_{ab}$ satisfy,

\begin{equation}
\frac {\partial A_{ab}(x(y))} {\partial y^{c}}
= \sum_{d=1}^{3}
\frac {\partial A_{ab}(x)} {\partial x^{d}}
 \frac {\partial x^{d}}{\partial y^{c}} 
\approx O(1/r^2), \quad \,
 \frac {\partial B_{ab}(y(x))} {\partial x^{c}}
= \sum_{d=1}^{3}
\frac {\partial B_{ab}(y)} {\partial y^{d}}
\frac {\partial y^{d}} {\partial x^{c}}
\approx O(1/r'^2).
\end{equation}
\\
Next, we consider the Christoffel symbols
corresponding to  $h_{(1)ab}$,$h_{(2)ab}$.
These take the asymptotic form,

\begin{equation}
\Gamma^{a}_{(1)bc} = \frac {\alpha}{r^2}
[\delta_{bc} x^{a} -
 \frac {x^{a} x_{b} x_{c}}{r^2}] + O(1/r^2),
\label{chris1}
\end{equation}
\\
and an analogous expression with
$(1) \leftrightarrow (2)$. The Christoffel
symbols $\Gamma^{a}_{(1)bc}$,
$\Gamma^{a}_{(2)bc}$ are related by,
\\
\begin{equation}
\Gamma^{a}_{(2)bc} =
\sum_{d,e,f=1}^{3}
\frac {\partial x^{d}}{\partial y^{b}}
\frac {\partial x^{e}}{\partial y^{c}}
\frac {\partial y^{a}}{\partial x^{f}}
\Gamma^{f}_{(1)de}
+ \sum_{f=1}^{3}
\frac {\partial^{2} x^{f}}{\partial y^{b} \partial y^{c}}
\frac {\partial y^{a}} {\partial x^{f}}.
\label{relation1}
\end{equation}
\\
Contracting eqs. (\ref{chris1}),
(\ref{relation1})
with $\delta_{ac}$ we obtain,
\\
\begin{equation}
\sum_{a=1}^{3} \Gamma^{a}_{(i)ba} =
 O(1/r^2),
\label{christo1}
\end{equation}

\begin{equation}
\sum_{a=1}^{3} \Gamma^{a}_{(2)ba} =
\sum_{d,f=1}^{3}
(\frac {\partial x^{d}}{\partial y^{b}}
\Gamma^{f}_{(1)df} +
\frac {\partial^{2} x^{f}}{\partial y^{b} \partial y^{d}}
\frac {\partial y^{d}} {\partial x^{f}}),
\label{relation3}
\end{equation}
\\
where $i = 1,2$.
From eqs. (\ref{acotado}),(\ref{christo1}),
(\ref{relation3}) we find,
\\
\begin{equation}
\sum_{a,f=1}^{3}
\frac {\partial^{2} x^{f}}{\partial y^{b} \partial y^{a}}
\frac {\partial y^{a}} {\partial x^{f}} =
O(1/r^2), \quad
\quad
\sum_{a,f=1}^{3}
\frac {\partial^{2} y^{f}} {\partial x^{b}\partial x^{a}}
\frac {\partial x^{a}} {\partial y^{f}} =
O(1/r^2).
\label{deri}
\end{equation}
\\
Eqs. (\ref{acotado}),(\ref{deri}) show that,
$\forall r>r_{0}$,
\\
\begin{equation}
\frac {\partial^{2} x^{a}} {\partial y^{c} \partial y^{b}}
= O(1/r^2),
\quad \quad
\frac {\partial^{2} y^{a}} {\partial x^{c} \partial x^{b}}
= O(1/r^2).
\label{derivada}
\end{equation}
\\
We integrate the eqs. (\ref{derivada})
on a curve $\Gamma$ on $\Sigma$,
with $\theta = \theta_{0}$,
$\varphi = \varphi_{0}$
($\theta_{0}$,$\varphi_{0} $ are constants), i.e.,
$\Gamma:[0,1] \mapsto \Sigma$,
$\Gamma(0) = r_{0}$, $\Gamma(1) = r$,
with $\Gamma(t) \geq r_{0}$ and  
unitary vector $n_{a}$ on $\Gamma$,
\\

\begin{equation}
\frac {\partial y^{a}} {\partial x^{b}}
= \sum_{c=1}^{3}
\int_{r_{0}}^{r}
\frac {\partial^2 y^{a}}{\partial x^{c}\partial x^{b}}
dx^{c} = \sum_{c=1}^{3} \int_{r_{0}}^{r}
\frac {\partial^2 y^{a}}{\partial x^{c}\partial x^{b}}
n^{c} dr = O(1/r)
+ W^{a}_{b} (\theta,\varphi),
\label{integral1}
\end{equation}
\\
\begin{equation}
\frac {\partial x^{a}} {\partial y^{b}}
= \sum_{c=1}^{3} \int_{r_{0}}^{r}
\frac {\partial^2 x^{a}}{\partial y^{c}\partial y^{b}}
dy^{c} = \sum_{c=1}^{3} \int_{r_{0}}^{r}
\frac {\partial^2 x^{a}}{\partial y^{c}\partial y^{b}}
n^{c} dr = O(1/r)
+ T^{a}_{b}(\theta,\varphi).
\label{integral2}
\end{equation}
\\
From these eqs., we conclude that,
\\
\begin{equation}
\lim_{r \to \infty}
\frac {\partial y^{a}} {\partial x^{b}}
= W^{a}_{b}, \quad \,
\lim_{r \to \infty}
\frac {\partial x^{a}} {\partial y^{b}}
= T^{a}_{b}.
\label{lim}
\end{equation}
\\
Noting that the only isometry of $h_{ab}^{0}$
is a rotation isometry, we can conclude
that $W^{a}_{b}, T^{a}_{b} \in SO(3)$ with
($W^{a}_{b}$, $T^{a}_{b}$) constant matrices.
From the properties,

\begin{equation}
\delta^{a}_{c} =
\lim_{r \to \infty} \sum_{b=1}^{3}
\frac {\partial y^{a}} {\partial x^{b}}
\frac {\partial x^{b}} {\partial y^{c}}
= \sum_{b=1}^{3} W^{a}_{b} T^{b}_{c},
\quad \quad \delta^{a}_{c} =
\lim_{r \to \infty} \sum_{b=1}^{3}
\frac {\partial x^{a}} {\partial y^{b}}
\frac {\partial y^{b}} {\partial x^{c}}
= \sum_{b=1}^{3} T^{a}_{b} W^{b}_{c},
\end{equation}
\\
we conclude $T^{a}_{b} = (W^{-1})^{a}_{b}$,
$W^{a}_{b} = (T^{-1})^{a}_{b}$.
Now, let's define,
\\
\begin{equation}
\eta^{a}(y) =
x^{a}(y) - \sum_{b=1}^{3} W^{a}_{b} y^{b},
\end{equation}

\begin{equation}
\zeta^{a}(x) =
y^{a}(x) - \sum_{b=1}^{3} T^{a}_{b} x^{b}.
\end{equation}
\\
Eqs. (\ref{integral1}),(\ref{integral2})
imply that,
\\
\begin{equation}
\frac {\partial \eta^{a}(y)} {\partial y^{b}}
\approx O(1/r),
\quad \,
\frac {\partial \zeta^{a}(x)} {\partial x^{b}}
\approx O(1/r).
\end{equation}
\\
Integrating these eqs., we have,

\begin{equation}
\eta^{a} \approx O(ln r), \quad \,
\zeta^{a} \approx O(ln r).
\label{ln}
\end{equation}
\\
Using the transformation (\ref{dif1})
in the eq. (\ref{Trans1}),
we obtain a new metric,

\begin{equation}
h_{(2)ab} = \sum_{c,e=1}^{3}
[(1-\alpha) \delta_{ce} +
\alpha \frac {x_{c}(y)x_{e}(y)} {r_{\eta}^{2}}
+ A_{ce}]
[\frac {\partial x^{c}(y)} {\partial y^{a}}]
[\frac {\partial x^{e}(y)} {\partial y^{b}}],
\label{h2}
\end{equation}

where,

\begin{equation}
x_{c}(y) = x^{c}(y) =
(\sum_{d=1}^{3} W^{c}_{d} y^{d} + \eta^{c}),
\end{equation}

\begin{equation}
\frac {\partial x^{c}(y)} {\partial y^{a}} =
W^{c}_{a} +
\frac {\partial \eta^{c}} {\partial y^{a}},
\end{equation}

\begin{equation}
r_{\eta}^{2} = \sum_{c=1}^{3}
(\sum_{d=1}^{3} (W^{c}_{d} y^{d})
+ \eta^{c})(\sum_{e=1}^{3}
(W^{c}_{e} y^{e})
+ \eta^{c}).
\end{equation}
\\
Using $r^2 = \sum_{e=1}^{3} y^{e}y^{e}$, 
we have the identity,

\begin{equation}
\frac {r_{\eta}^{2}}{r^2} =
1 + 2 \sum_{c,e=1}^{3} W^{c}_{e}
\frac {y^{e} \eta^{c}}{r^2} +
\sum_{e=1}^{3}
\frac {\eta^{e} \eta^{e}}{r^2},
\end{equation}

With $\eta^{a} \approx O(ln r)$,
we expand the above identity,

\begin{equation}
(\frac {r_{\eta}^{2}}{r^2})^{-1} =
1 - 2 \sum_{c,e=1}^{3} W^{c}_{e}
\frac {y^{e} \eta^{c}}{r^2} + ...
= 1 + O(\frac {ln r}{r}).
\label{exp-r-eta}
\end{equation}

Introducing the eq. (\ref{exp-r-eta})
in the eq. (\ref{h2}) we find,
\\
\begin{eqnarray}
h_{(2)ab} = (1-\alpha) \delta_{ab}
+ \frac {\alpha}{r^2} y_{a}y_{b}
+ \frac {\alpha}{r^2}
\sum_{e=1}^{3} \eta_{e}
[ W^{e}_{a} y_{b} + W^{e}_{b} y_{a}
- \frac {2}{r^2} \sum_{f=1}^{3}
W^{e}_{f} y^{f} y_{a} y_{b} ]
+ O(\frac {\eta^2}{r^2}).
\nonumber \\
\nonumber \\
\label{expmetr}
\end{eqnarray}
\\
Then the function $\eta^{a} \approx O(ln r)$
does not preserve the asymptotic
form of the metric $h_{(1)ab}$,
eq. (\ref{met1}). However, we can see
from the above expansion of
$h_{(2)ab}$ in (\ref{expmetr}) that
we need $\eta^{a} \approx O(1)$, in order
to preserve the asymptotic
form of the metric $h_{(1)ab}$, which 
establishes the lemma.
\\
\\

{\bf Theorem}: The value of ADM$\alpha$
mass is independent of diffeomorphisms
$\Phi$ that preserve the asymptotic
form of the metric ${\bf h}$.
\\
\\
Proof: Let's write the metric ${\bf h}$
in a coordinate system $\{x^{a}\}$,
eq. (\ref{met1}),
\\
\begin{equation}
h_{(1)ab} = (1-\alpha) \delta_{ab}
+ \alpha x_a x_b/r^2 + A_{ab},
\label{met-asim}
\end{equation}
\\
and we take a diffeomorphism, $\Phi$, that
preserves the asymptotic form of the
metric ${\bf h}$,
eq. (\ref{dif1}),
\\
\begin{equation}
x_{a}(y) = x^{a}(y) = \sum_{b=1}^{3}
W^{a}_{b} y^{b} + \eta^{a}(y),
\label{difeomorfismo}
\end{equation}
\\
where,
\\
\begin{equation}
A_{ab} \approx O(1/r), \quad \,
\frac {\partial A_{ab}} {\partial x^{c}}
\approx O(1/r^2), \quad \,
\eta^{a} \approx O(1), \quad \,
\frac {\partial \eta^{a}} {\partial y^{b}}
\approx O(1/r).
\label{condi3}
\end{equation}
\\
The value of the ADM$\alpha$ mass for
the metric $h_{(1)ab}$, from
eq. (\ref{masa}), is

\begin{eqnarray}
16 \pi (1-\alpha) M^{(1)}_{ADM\alpha}
& = & \int_{\partial\Sigma} dS_{a}^{(x)}
( h^{0ac}_{(x)} h^{0bd}_{(x)} 
- h^{0ab}_{(x)} h^{0cd}_{(x)} )
\, D_{b}^0 (h_{(1)cd})
\nonumber \\
& = & \int_{\partial\Sigma}
dS_{a}^{(x)} h^{0ac}_{(x)} h^{0bd}_{(x)} \,
[ \frac {\partial h_{(1)cd}} {\partial x^{b}}
- \frac {\partial h_{(1)bd}} {\partial x^{c}}
+ \Gamma^{e(0)}_{cd} h_{(1)eb}
- \Gamma^{e(0)}_{bd} h_{(1)ec} ], 
\nonumber \\
\label{masaADM}
\end{eqnarray}

where,

\begin{equation}
h^{0bd}_{(x)} = \frac{1}{(1 - \alpha)}
(\delta^{bd} -
\frac {\alpha} {r^2} x^{b}x^{d}),
\label{met-inv}
\end{equation}
\\
$dS_{a}^{(x)}$ and $\Gamma^{e(0)}_{cd}$ 
are the surface element and the
Christoffel symbols
related with the  S.A.F.D.A $\alpha$ spatial
metric $h_{ac(x)}^{0}$,
\\
\begin{equation}
\Gamma^{e(0)}_{cd} = \frac {\alpha} {r^2}
(\delta_{cd} x^{e} -
\frac {x^{e}x_{c}x_{d}} {r^2}).
\label{chris-standard}
\end{equation}
\\
Introducing eqs. (\ref{met-inv}) and
(\ref{chris-standard}) in the
eq. (\ref{masaADM}),
we obtain,
\\
\begin{equation}
16 \pi (1-\alpha) M^{(1)}_{ADM\alpha} =
\int_{\partial\Sigma} \, \, dS_{a}^{(x)} \,
( K_{(1)}^{a} + F_{(1)}^{a} ),
\label{masa1}
\end{equation}
\\
where $K_{(1)}^{a}$, $F_{(1)}^{a}$,
are given by:
\\
\begin{equation}
K_{(1)}^{a} = h^{0ac}_{(x)} h^{0bd}_{(x)} \,
[ \frac {\partial h_{(1)cd}} {\partial x^{b}}
- \frac {\partial h_{(1)bd}} {\partial x^{c}} ],
\end{equation}
\\
\begin{equation}
F_{(1)}^{a} =
\frac {\alpha}{(1 - \alpha)^{2} r^2} \,
[ - \delta^{ac} x^{b} h_{(1)bc}
+ \frac {(2\alpha - 1)} {r^2}
x^{a} x^{b} x^{c} h_{(1)bc} ].
\end{equation}
\\
We introduce the metric $h_{(1)ab}$,
eq. (\ref{met-asim}), in the eq. (\ref{masa1}),
and eliminating terms with vanishing contribution
to the integral, we obtain,
\\
\begin{equation}
16 \pi (1-\alpha) M^{(1)}_{ADM\alpha} =
\int_{\partial\Sigma}
dS_{a}^{(x)} \, ( R_{(1)}^{a} + J_{(1)}^{a} ),
\label{masa1-asim}
\end{equation}
\\
where $R_{(1)}^{a}$, $J_{(1)}^{a}$, are
given by:
\\
\begin{equation}
R_{(1)}^{a} = h^{0ac}_{(x)} h^{0bd}_{(x)}
[ \frac {\partial A_{cd}} {\partial x^{b}}
- \frac {\partial A_{bd}} {\partial x^{c}} ],
\end{equation}
\\
\begin{equation}
J_{(1)}^{a} =
\frac {\alpha}{(1 - \alpha)^{2} r^2}
[- \sum_{b,c=1}^{3} \delta^{ac} x^{b} A_{bc}
+ \frac {(2\alpha - 1)} {r^2}
\sum_{b,c=1}^{3} x^{a} x^{b} x^{c} A_{bc}].
\end{equation}
\\
We introduce the transformation (\ref{difeomorfismo})
in the eq. (\ref{Trans1}), and we obtain
the new form of the metric,
\\
\begin{equation}
h_{(2)ab} = \sum_{c,e=1}^{3}
[(1-\alpha) \delta_{ce} +
\alpha \frac {x_{c}(y)x_{e}(y)} {r_{\eta}^{2}}
+ A_{ce}]
[\frac {\partial x^{c}(y)} {\partial y^{a}}]
[\frac {\partial x^{e}(y)} {\partial y^{b}}],
\label{metrica3}
\end{equation}
\\
where,

\begin{equation}
x_{c}(y) = x^{c}(y) =
(\sum_{d=1}^{3} W^{c}_{d} y^{d} + \eta^{c}),
\end{equation}

\begin{equation}
\frac {\partial x^{c}(y)} {\partial y^{a}} =
W^{c}_{a}
+ \frac {\partial \eta^{c}} {\partial y^{a}}
\label{x-parcial-y}
\end{equation}

\begin{equation}
r_{\eta}^{2} = \sum_{c=1}^{3}
(\sum_{d=1}^{3} (W^{c}_{d} y^{d}) + \eta^{c})
(\sum_{c=1}^{3} (W^{c}_{e} y^{e}) + \eta^{c}).
\label{r-eta}
\end{equation}
\\
Now, we calculate the ADM$\alpha$ mass for
the metric $h_{(2)ab}$,
we obtain,

\begin{eqnarray}
16 \pi (1-\alpha) M^{(2)}_{ADM\alpha} & = &
\int_{\partial\Sigma}
dS_{a}^{(y)} ( h^{0ac}_{(y)} h^{0bd}_{(y)} 
             - h^{0ab}_{(y)} h^{0cd}_{(y)} )
D_{b}^0 (h_{(2)cd})
\nonumber \\
\nonumber \\
& = &
\int_{\partial\Sigma}
dS_{a}^{(y)} \, ( K_{(2)}^{a} + F_{(2)}^{a} ),
\label{masa2}
\end{eqnarray}
\\
where $K_{(2)}^{a}$, $F_{(2)}^{a}$, are:
\\
\begin{equation}
K_{(2)}^{a} = h^{0ac}_{(y)} h^{0bd}_{(y)} \,
[ \frac {\partial h_{(2)cd}} {\partial y^{b}}
- \frac {\partial h_{(2)bd}} {\partial y^{c}} ],
\label{expresion1}
\end{equation}
\\
\begin{equation}
F_{(2)}^{a} =
\frac {\alpha}{(1 - \alpha)^{2} r^2} \,
[ - \delta^{ac} y^{b} h_{(2)bc}
+ \frac {(2\alpha - 1)} {r^2}
y^{a} y^{b} y^{c} h_{(2)bc} ],
\label{expresion2}
\end{equation}
\\
$dS_{a}^{(y)}$ is the surface 
element associated with 
spatial metric $h_{bd(y)}^{0}$ and,
\\
\begin{equation}
h^{0bd}_{(y)} = \frac{1}{(1 - \alpha)}
(\delta^{bd} -
\frac {\alpha} {r^2} y^{b}y^{d}).
\end{equation}
\\
From eq. (\ref{metrica3}), we have,
\begin{eqnarray}
\frac {\partial h_{(2)cd}} {\partial y^{b}} & = &
\sum_{e,f=1}^{3}
[ \frac {\alpha} {r_{\eta}^2}
( x_{e} \frac {\partial x_{f}} {\partial y^{b}} +
  x_{f} \frac {\partial x_{e}} {\partial y^{b}} -
\frac {2} {r_{\eta}^2}
\sum_{a=1}^{3} x_{e} x_{f} x_{a}
\frac {\partial x_{a}} {\partial y^{b}} ) +
\frac {\partial A_{ef}} {\partial y^{b}} ] \,
\frac {\partial x^{e}} {\partial y^{c}}   
\frac {\partial x^{f}} {\partial y^{d}}
\nonumber \\
\nonumber \\
& + & \sum_{e,f=1}^{3}
[ (1-\alpha) \delta_{ef}
+ \frac {\alpha} {r_{\eta}^2} x_{e} x_{f}
+ A_{ef} ] \,
[\frac{\partial^{2} x^{e}}{\partial y^{b}\partial y^{c}}
\frac {\partial x^{f}} {\partial y^{d}} +
\frac {\partial^{2} x^{f}}{\partial y^{b}\partial y^{d}} 
\frac {\partial x^{e}} {\partial y^{c}} ].
\label{dermetrica3}
\end{eqnarray}
\\
Introducing the eq. (\ref{dermetrica3}) in
the eq. (\ref{expresion1}), we obtain,
\\
\begin{equation}
\int_{\partial\Sigma}
dS_{a}^{(y)} K_{(2)}^{a} =
\int_{\partial\Sigma}
dS_{a}^{(y)} ( Q^{a} + P_{(1)}^{a} + H^{a}),
\label{masa2-(1)}
\end{equation}
\\
where,

\begin{equation}
Q^{a} = h^{0ac}_{(y)} h^{0bd}_{(y)} 
\, \sum_{e,f=1}^{3}
[ \frac {\partial A_{ef}} {\partial y^{b}} 
\frac {\partial x^{e}} {\partial y^{c}}   
\frac {\partial x^{f}} {\partial y^{d}} -
\frac {\partial A_{ef}} {\partial y^{c}} 
\frac {\partial x^{e}} {\partial y^{d}}   
\frac {\partial x^{f}} {\partial y^{b}} ],
\label{term(1)}
\end{equation}
\\
\begin{equation}
P_{(1)}^{a} = h^{0ac}_{(y)} h^{0bd}_{(y)} \,
(\frac {\alpha} {r_{\eta}^2}) \, \sum_{e,f=1}^{3}
[ x_{e} \frac {\partial x_{f}} {\partial y^{b}} 
\frac {\partial x^{e}} {\partial y^{c}}   
\frac {\partial x^{f}} {\partial y^{d}} -
  x_{f} \frac {\partial x_{e}} {\partial y^{c}} 
\frac {\partial x^{e}} {\partial y^{d}}   
\frac {\partial x^{f}} {\partial y^{b}} ],
\label{term(2)}
\end{equation}
\\
\begin{equation}
H^{a} = h^{0ac}_{(y)} h^{0bd}_{(y)}
\sum_{e,f=1}^{3} [ (1-\alpha) \delta_{ef} +
\frac {\alpha} {r_{\eta}^2} x_{e} x_{f} + A_{ef} ] \,
[ \frac {\partial^{2} x^{f}} {\partial y^{b} \partial y^{d}}
\frac {\partial x^{e}} {\partial y^{c}} -
\frac {\partial^{2} x^{e}} {\partial y^{c} \partial y^{d}} 
\frac {\partial x^{f}} {\partial y^{b}} ].
\label{term(3)}
\end{equation}
\\
Introducing the eq. (\ref{metrica3}) in the
eq. (\ref{expresion2}), we obtain,
\\
\begin{equation}
\int_{\partial\Sigma}
\sum_{a=1}^{3} dS_{a}^{(y)} F_{(2)}^{a} =
\int_{\partial\Sigma}
\frac {\alpha}{(1 - \alpha)^{2} r^2}
\sum_{a=1}^{3} dS_{a}^{(y)}  \,
( P_{(2)}^{a} + P_{(3)}^{a} + V^{a} ),
\label{masa2-(2)}
\end{equation}
\\
where,

\begin{equation}
P_{(2)}^{a} =
(1 - \alpha) \sum_{e,f,b,c =1}^{3} \delta_{ef} \,
[ - \delta^{ac} y^{b}
\frac {\partial x^{e}} {\partial y^{c}} 
\frac {\partial x^{f}} {\partial y^{b}}
+ \frac {(2\alpha - 1)} {r^2} y^{a} y^{b} y^{c}
\frac {\partial x^{e}} {\partial y^{c}}
\frac {\partial x^{f}} {\partial y^{b}} ],
\label{term(4)}
\end{equation}
\\
\begin{equation}
P_{(3)}^{a} =
(\frac {\alpha}{r_{\eta}^2})
\sum_{e,f,b,c=1}^{3} x_{e} x_{f} \,
[ - \delta^{ac} y^{b}
\frac {\partial x^{e}} {\partial y^{c}} 
\frac {\partial x^{f}} {\partial y^{b}}
+ \frac {(2\alpha - 1)} {r^2} y^{a} y^{b} y^{c}
\frac {\partial x^{e}} {\partial y^{c}}
\frac {\partial x^{f}} {\partial y^{b}} ],
\label{term(5)}
\end{equation}
\\
\begin{equation}
V^{a} = \sum_{e,f,b,c=1}^{3} A_{ef} \,
[ - \delta^{ac} y^{b}
\frac {\partial x^{e}} {\partial y^{c}} 
\frac {\partial x^{f}} {\partial y^{b}}
+ \frac {(2\alpha - 1)} {r^2} y^{a} y^{b} y^{c}
\frac {\partial x^{e}} {\partial y^{c}}
\frac {\partial x^{f}} {\partial y^{b}} ].
\label{term(6)}
\end{equation}
\\
The different expressions have been
named according to the way in which the terms
will combine in the subsequent evaluation.
We will see that the expressions
$Q^{a}$ and $V^{a}$ reproduce
the $M^{(1)}_{ADM\alpha}$, and that the
integral of the remainder terms is null.
From these last terms, first, we will
concentrate on $P_{(1)}^{a}$, $P_{(2)}^{a}$,
$P_{(3)}^{a}$. Using the expansion
(\ref{exp-r-eta}),
the eqs. (\ref{difeomorfismo}),
(\ref{x-parcial-y}),
and the asymptotic conditions (\ref{condi3}), 
and eliminating terms with vanishing
contribution at
the boundary $\partial\Sigma$, we obtain,

\begin{eqnarray}
\int_{\partial\Sigma} 
dS_{a}^{(y)} P_{(1)}^{a} & = & 
\int_{\partial\Sigma}
\frac {\alpha} {(1 - \alpha)^{2} r^2}
\sum_{a=1}^{3} dS_{a}^{(y)} \,
[ \, 2(1 - \alpha) y^{a}
+ (2 - \alpha) \sum_{e,f=1}^{3}
\delta^{af} W^{e}_{f} \eta_{e}
\nonumber \\
\nonumber \\
&& \quad
- (4 - 3\alpha) \sum_{e,f=1}^{3}
W^{e}_{f} \eta_{e} \frac {y^{f} y^{a}} {r^2}
+ 2(1 - \alpha) \sum_{e,f=1}^{3}
W^{e}_{f}
\frac {\partial \eta_{e}} {\partial y^{f}}
y^{a}
\nonumber \\
\nonumber \\
&& \quad
+ \sum_{c,e,f=1}^{3}
\delta^{ac} W^{e}_{f}
\frac {\partial \eta_{e}} {\partial y^{c}}
y^{f}
+ (\alpha - 1) \sum_{c,e,f=1}^{3}
\delta^{ac} W^{e}_{c}
\frac {\partial \eta_{e}} {\partial y^{f}}
y^{f}
\nonumber \\
\nonumber \\
&& \quad
- (\frac {\alpha} {r^2})
\sum_{d,e,f=1}^{3} W^{e}_{f}
\frac {\partial \eta_{e}} {\partial y^{d}}
y^{f} y^{d} y^{a} \, ],
\label{term(9)}
\end{eqnarray}
\\
in the same way, we obtain,
\\
\begin{eqnarray}
\int_{\partial\Sigma}
\frac {1}{r^2}
\sum_{a=1}^{3} dS_{a}^{(y)} P_{(2)}^{a} & = &
\int_{\partial\Sigma}
\frac {(1 - \alpha)}{r^2}
\sum_{a=1}^{3} dS_{a}^{(y)} \,
[ \, 2(\alpha - 1) y^{a}
- \sum_{c,e,f=1}^{3} \delta^{ac} W^{e}_{f}
\frac {\partial \eta_{e}} {\partial y^{c}}
y^{f}
\nonumber \\
\nonumber \\
&& \quad
- \sum_{c,e,f=1}^{3} \delta^{ac} W^{e}_{c}
\frac {\partial \eta_{e}} {\partial y^{f}}
y^{f}
+ \frac {(4\alpha - 2)} {r^2}
\sum_{c,e,f=1}^{3} W^{e}_{c}
\frac {\partial \eta_{e}} {\partial y^{f}}
y^{f} y^{c} y^{a} \, ],
\nonumber \\
\nonumber \\
\label{term(11)}
\end{eqnarray}
\\
and,
\\
\begin{eqnarray}
\int_{\partial\Sigma} \sum_{a=1}^{3}
\frac {dS_{a}^{(y)}}{r^2}
P_{(3)}^{a} & = &
\int_{\partial\Sigma}
\frac {\alpha}{r^2}
\sum_{a=1}^{3} dS_{a}^{(y)} \,
[ 2(\alpha - 1) y^{a}
- \sum_{c,e=1}^{3} \delta^{ac} W^{e}_{c} \eta_{e}
- \sum_{c,e,f=1}^{3} \delta^{ac} W^{e}_{f}
\frac {\partial \eta_{e}} {\partial y^{c}} y^{f} 
\nonumber \\
\nonumber \\
&& \quad
+ \frac {(4\alpha - 3)} {r^2}
\sum_{c,e,f=1}^{3} W^{e}_{c}
\frac {\partial \eta_{e}} {\partial y^{f}}
y^{f} y^{c} y^{a}
+ \sum_{e,f=1}^{3} W^{e}_{f} \eta_{e}
\frac {y^{f} y^{a}} {r^2} ].
\nonumber \\
\nonumber \\
\label{term(12)}
\end{eqnarray}
\\
Adding eqs. (\ref{term(9)}),
(\ref{term(11)}), (\ref{term(12)}),
we obtain,
\\
\begin{eqnarray}
\int_{\partial\Sigma}
\sum_{a=1}^{3} dS_{a}^{(y)} P^{a} &= &
\int_{\partial\Sigma}
\frac {\alpha} {(1 - \alpha) r^2}
\sum_{a,e,f=1}^{3} dS_{a}^{(y)} \,
[ \, 2 \delta^{af} W^{e}_{f} \eta_{e}
- 4 W^{e}_{f} \eta_{e} \frac {y^{f} y^{a}} {r^2}
\nonumber \\
\nonumber \\
&&
+ 2 \sum_{c=1}^{3} \delta^{cf} W^{e}_{f}
\frac {\partial \eta_{e}} {\partial y^{c}} y^{a}
- 2 \sum_{c=1}^{3} \delta^{af} W^{e}_{f}
\frac {\partial \eta_{e}} {\partial y^{c}} y^{c}
- 2 \sum_{c=1}^{3} W^{e}_{f}
\frac {\partial \eta_{e}} {\partial y^{c}}
\frac {y^{c} y^{f} y^{a}} {r^2} \, ], 
\nonumber \\
\nonumber \\
\label{term(14)}
\end{eqnarray}
\\
where
$P^{a} = P_{(1)}^{a} + P_{(2)}^{a} + P_{(3)}^{a}$.
Now, we will concentrate on $H^{a}$.
Again, using eqs. (\ref{difeomorfismo}),
(\ref{x-parcial-y}) and the asymptotic
conditions (\ref{condi3}),
and eliminating terms with vanishing
contribution, we find,
\\
\begin{equation}
\int_{\partial\Sigma} dS_{a}^{(y)} H^{a} =
\int_{\partial\Sigma} \sum_{a=1}^{3}
dS_{a}^{(y)} \,
( H_{(1)}^{a} + H_{(2)}^{a} ),
\label{term(10)}
\end{equation}
\\
where,

\begin{equation}
\int_{\partial\Sigma} \sum_{a=1}^{3}
dS_{a}^{(y)} H_{(1)}^{a} =
\int_{\partial\Sigma} \frac {1} {(1 - \alpha)}
\sum_{a,c,e,f=1}^{3} dS_{a}^{(y)} \, \delta^{af} \,
[ W^{e}_{f}
\frac {\partial^{2} \eta_{e}}{\partial y^{c} \partial y_{c}}
- W^{e}_{c}
\frac {\partial^{2} \eta_{e}}{\partial y^{f} \partial y_{c}} ],
\label{term(10-1)}
\end{equation}
\\
\begin{eqnarray}
\int_{\partial\Sigma} \sum_{a=1}^{3}
dS_{a}^{(y)} H_{(2)}^{a} =
\int_{\partial\Sigma} 
\frac {\alpha \, \, dS_{a}^{(y)}} {(1 - \alpha) r^2}
\sum_{a,c,d,e,f=1}^{3} \delta^{ad} \,
[ W^{e}_{c}
\frac{\partial^{2}\eta_{e}}{\partial y^{f}\partial y_{c}}
y^{f} y_{d} -
  W^{e}_{d}
\frac{\partial^{2}\eta_{e}}{\partial y_{c}\partial y^{f}}
y^{f} y_{c} ].
\nonumber \\
\nonumber \\
\label{term(10-2)}
\end{eqnarray}
\\
We will see that $H_{(1)}^{a}$ vanishes by itself.
Using the eq. (\ref{term(10-1)}), we find,
\\
\begin{eqnarray}
\int_{\partial\Sigma} \sum_{a=1}^{3}
dS_{a}^{(y)} H_{(1)}^{a} & = &
\frac {1} {(1 - \alpha)}
\int_{\partial\Sigma} \sum_{a,c,e,f=1}^{3}
dS_{a}^{(y)} \, \delta^{af} \,
 [ \, W^{e}_{f} \frac {\partial^{2} \eta_{e}}
{\partial y^{c} \partial y_{c}} -
  W^{e}_{c} \frac {\partial^{2} \eta_{e}}
{\partial y^{f} \partial y_{c}} \, ]
\nonumber \\
\nonumber \\
& = & \frac {1} {(1 - \alpha)}
\int_{\partial\Sigma} \sum_{a,c=1}^{3} 
dS_{a}^{(y)}
\frac {\partial} {\partial y_{c}}
(\gamma^{a}_{c})
\nonumber \\
\nonumber \\
& = & 
\frac {1} {(1 - \alpha)}
\int_{\Sigma} \sum_{a,c=1}^{3} 
\frac {\partial^2} {\partial y^{a} \partial y_{c}}
(\gamma^{a}_{c}) = 0,
\label{term(15)}
\end{eqnarray}
\\
the last equality follows from the fact that
$\gamma^{a}_{c}$ is an antisymmetric quantity,
\\
\begin{equation}
\gamma^{a}_{c} =
\sum_{e,f=1}^{3} \, \delta^{af} \,
[ \, W^{e}_{f}
\frac {\partial \eta_{e}} {\partial y^{c}}
   - W^{e}_{c}
\frac {\partial \eta_{e}} {\partial y^{f}} \, ]. 
\end{equation}
\\
We integrate by parts the 
eq. (\ref{term(10-2)}),

\begin{eqnarray}
\int_{\partial\Sigma}
\sum_{a=1}^{3} dS_{a}^{(y)}
H_{(2)}^{a} & = &
\frac {\alpha} {(1 - \alpha)}
\int_{\partial\Sigma}
\sum_{a,c,d,e,f=1}^{3} dS_{a}^{(y)}
\, \delta^{ad} \,
\frac {\partial} {\partial y_{c}} \,
[ \, W^{e}_{c}
\frac {\partial \eta_{e}} {\partial y^{f}}
  \frac {y^{f} y_{d}} {r^2} -
     W^{e}_{d}
\frac {\partial \eta_{e}} {\partial y^{f}}
  \frac {y^{f} y_{c}} {r^2} \, ]
\nonumber \\
\nonumber \\
& + & \frac {\alpha} {(1 - \alpha)}
\int_{\partial\Sigma}
\sum_{a,c,d,e,f=1}^{3} dS_{a}^{(y)} 
\, \delta^{ad} \,
(\frac {\partial \eta_{e}} {\partial y^{f}}) \,
 \frac {\partial} {\partial y_{c}}
[ W^{e}_{d} \frac {y^{f} y_{c}} {r^2} -
  W^{e}_{c} \frac {y^{f} y_{d}} {r^2} ].
\nonumber \\
\nonumber \\
\label{term(16)}
\end{eqnarray}
\\
The first surface integral of the
eq. (\ref{term(16)}) is the integral of
the divergence of an antisymmetric quantity, 
and, therefore, it is null
(as in equation (\ref{term(15)})).
Next, using the identities,
\\
\begin{equation}
\sum_{c,d,e,f=1}^{3} \, \delta^{ad} \,
(\frac {\partial \eta_{e}} {\partial y^{f}})
\frac {\partial} {\partial y_{c}}
(W^{e}_{d} \frac {y^{f} y_{c}} {r^2}) =
\sum_{d,e,f=1}^{3} \,
2 \delta^{ad} \,
(\frac {\partial \eta_{e}} {\partial y^{f}})
W^{e}_{d} \frac {y^{f}} {r^2},
\label{term(18)}
\end{equation}
\\
\begin{eqnarray}
\sum_{c,d,e,f=1}^{3} \, \delta^{ad} \,
(\frac {\partial \eta_{e}} {\partial y^{f}})
\frac {\partial} {\partial y_{c}}
(W^{e}_{c} \frac {y^{f} y_{d}} {r^2}) & = &
\sum_{d,e,f=1}^{3} \, \delta^{ad} \,
[ (\frac {\partial \eta_{e}} {\partial y^{f}})
W^{e}_{d} \frac {y^{f}} {r^2} +
(\frac {\partial \eta_{e}} {\partial y^{f}})
W^{e}_{f} \frac {y_{d}} {r^2} ]
\nonumber \\
\nonumber \\
&& - 2 \sum_{c,d,e,f=1}^{3} \, \delta^{ad} \,
(\frac {\partial \eta_{e}} {\partial y^{f}})
W^{e}_{c} \frac {y_{d} y^{f} y^{c}} {r^4},
\nonumber \\
\nonumber \\
\label{term(19)}
\end{eqnarray}
\\
in the eq. (\ref{term(16)}), we obtain,

\begin{eqnarray}
\int_{\partial\Sigma} \sum_{a=1}^{3}
dS_{a}^{(y)} H_{(2)}^{a} & = &
\frac {\alpha} {(1 - \alpha)}
\int_{\partial\Sigma}
\sum_{a,c,e,f=1}^{3} dS_{a}^{(y)} \,
[ \, 2 \, \delta^{ac} \,
(\frac {\partial \eta_{e}} {\partial y^{f}})
W^{e}_{c} \frac {y^{f}} {r^2}
      - \delta^{ac} \,
(\frac {\partial \eta_{e}} {\partial y^{f}})
W^{e}_{c} \frac {y^{f}} {r^2}
\nonumber \\
\nonumber \\
&& \quad \quad \quad \quad
      - \delta^{ac} \,
  (\frac {\partial \eta_{e}} {\partial y^{f}})
W^{e}_{f} \frac {y_{c}} {r^2} +
2 (\frac {\partial \eta_{e}} {\partial y^{f}})
W^{e}_{c} \frac {y^{a} y^{f} y^{c}} {r^4} \, ].
\label{term(20)}
\end{eqnarray}
\\
Adding eqs. (\ref{term(14)}), (\ref{term(15)})
and (\ref{term(20)}), we find for the
$M^{(2)}_{ADM\alpha}$, eq. (\ref{masa2})),
\\
\begin{eqnarray}
16 \pi (1-\alpha) M^{(2)}_{ADM\alpha}
& = & \int_{\partial\Sigma}
\sum_{a=1}^{3} dS_{a}^{(y)} \,
[ \, Q^{a} + P^{a} + H_{(2)}^{a}
+ \frac {\alpha}{(1 - \alpha)^{2} r^2} V^{a} \, ],
\nonumber \\
\label{masatotal}
\end{eqnarray}
\\
where,
\\
\begin{eqnarray}
\int_{\partial\Sigma} 
\sum_{a=1}^{3} dS_{a}^{(y)}
[ P^{a} + H_{(2)}^{a} ] & = &
\frac {\alpha} {(1 - \alpha)}
\int_{\partial\Sigma}
\sum_{a,e,f=1}^{3} dS_{a}^{(y)} 
(\frac {1} {r^2}) \,
[ \, 2 \delta^{af} \,
W^{e}_{f} \eta_{e}
- 4 W^{e}_{f} \eta_{e} 
\frac {y^{f} y^{a}} {r^2}
\nonumber \\
\nonumber \\
&& \quad \quad \quad \quad
+ \sum_{c=1}^{3} \delta^{ac} \,
\{
  (\frac {\partial \eta_{e}} {\partial y_{f}})
W^{e}_{f} y_{c}
- (\frac {\partial \eta_{e}} {\partial y_{f}})
W^{e}_{c} y_{f} \, \} \, ].
\nonumber \\
\nonumber \\
\label{term(21)}
\end{eqnarray}
\\
We will see that the above integral is vanishing.
We integrate by parts the last two terms
of the integral (\ref{term(21)}),
\\
\begin{eqnarray}
\int_{\partial\Sigma}
\sum_{a,c,e,f=1}^{3}
\frac {dS_{a}^{(y)}} {r^2} \delta^{ac} \,
\frac {\partial \eta_{e}} {\partial y_{f}}
[ W^{e}_{f} y_{c} - W^{e}_{c} y_{f} ] & = &
\int_{\partial\Sigma}
\sum_{a,c,e,f=1}^{3} 
dS_{a}^{(y)} \delta^{ac} \,
\frac {\partial} {\partial y_{f}}
[ \eta_{e} \frac {W^{e}_{f} y_{c}} {r^2} -
  \eta_{e} \frac {W^{e}_{c} y_{f}} {r^2} ]
\nonumber \\
\nonumber \\
& + & \int_{\partial\Sigma}
\sum_{a,c,e,f=1}^{3} dS_{a}^{(y)} 
\delta^{ac} \,
\eta_{e} \frac {\partial} {\partial y_{f}}
[ \frac {W^{e}_{c} y_{f}} {r^2} -
  \frac {W^{e}_{f} y_{c}} {r^2} ].
\nonumber \\
\nonumber \\
\label{term(22)}
\end{eqnarray}
\\
The first surface integral of the
eq. (\ref{term(22)})
is the integral of the divergence
of an antisymmetric quantity, and,
therefore, it is null
(as in equation (\ref{term(15)})).
Now making use of the identity,
\\
\begin{equation}
\sum_{c,e,f=1}^{3} \, \delta^{ac} \,
[ \, \eta_{e} W^{e}_{c}
\frac {\partial} {\partial y_{f}}
( \frac {y_{f}} {r^2} ) -
     \eta_{e} W^{e}_{f}
\frac {\partial} {\partial y_{f}}
( \frac {y_{c}} {r^2} ) \, ] =
2 \sum_{e,f=1}^{3}
\eta_{e} W^{e}_{f} \frac {y^{f} y^{a}} {r^4},
\end{equation}
\\
\\
in the eq. (\ref{term(22)}), we obtain,
\\
\\
\begin{equation}
\int_{\partial\Sigma}
\sum_{a,c,e,f=1}^{3}
\frac {dS_{a}^{(y)}} {r^2} \, \delta^{ac} \,
(\frac {\partial \eta_{e}} {\partial y_{f}})
[ \, W^{e}_{f} y_{c}
   - W^{e}_{c} y_{f} \, ] =
\int_{\partial\Sigma}
\sum_{a,e,f=1}^{3} dS_{a}^{(y)} \,
[ \, 2 \eta_{e} W^{e}_{f}
\frac {y^{f} y^{a}} {r^4} \, ].
\label{term(24)}
\end{equation}
\\
\\
Substituting (\ref{term(24)})
in (\ref{term(21)}), we find,
\\
\\
\begin{equation}
\int_{\partial\Sigma} \sum_{a=1}^{3}
dS_{a}^{(y)} \, [ P^{a} + H_{(2)}^{a} ] =
\frac {\alpha} {(1 - \alpha)}
\int_{\partial\Sigma}
\sum_{a,e,f=1}^{3} 
\frac {dS_{a}^{(y)}} {r^2} \,
[ \, 2 \, \delta^{af} \, W^{e}_{f} \eta_{e}
   - 2 W^{e}_{f} \eta_{e}
\frac {y^{f} y^{a}} {r^2} \, ].
\nonumber \\
\label{term(25)}
\end{equation}
\\
\\
We calculate the above integrals using a new
rotated quantity,
\\
\\
\begin{equation}
\overline{\eta_{f}} = 
\sum_{e=1}^{3} W^{e}_{f} \eta_{e},
\end{equation}
\\
then we have, for (\ref{term(25)}),
\\
\begin{equation}
\int_{\partial\Sigma} \sum_{a=1}^{3}
dS_{a}^{(y)} \, [ P^{a} + H_{(2)}^{a} ] =
\frac {\alpha} {(1 - \alpha)}
\int_{\partial\Sigma}
\sum_{a=1}^{3} \frac {dS_{a}} {r^2} \,
[ \, 2 \overline{\eta_{a}} - 
2 \sum_{e=1}^{3} \overline{\eta_{e}} 
\frac{y^{e} y^{a}} {r^2} \, ].
\label{term(26)}
\end{equation}
\\
Using the properties
\\
\begin{equation}
\sum_{a=1}^{3} dS_{a}^{(y)}
\overline{\eta_{a}} = 
dS_{r}^{(y)} \overline{\eta_{r}}, 
\quad
\sum_{a=1}^{3} dS_{a}^{(y)}
\frac {y^{a}} {r} = 
dS_{r}^{(y)}, \quad
\sum_{a=1}^{3} \overline{\eta_{a}}
\frac {y^{a}} {r} = 
\overline{\eta_{r}}, \quad 
\label{term(27)}
\end{equation}
\\
in (\ref{term(26)}), we obtain,
\\
\begin{equation}
\int_{\partial\Sigma} \sum_{a=1}^{3}
dS_{a}^{(y)} \, 
[ P^{a} + H_{(2)}^{a} ] = 0.
\label{term(28)}
\end{equation}
\\
Introducing eq.
(\ref{term(28)}) in 
(\ref{masatotal}), we have,
\\
\begin{equation}
16 \pi (1-\alpha) M^{(2)}_{ADM\alpha} =
\int_{\partial\Sigma}
\sum_{a=1}^{3} dS_{a}^{(y)} \,
[ \, Q^{a}
+ \frac {\alpha}{(1 - \alpha)^{2} r^2}
V^{a} \, ],
\label{masatotal(2)}
\end{equation}
\\
Using 
the eqs. (\ref{difeomorfismo}),
(\ref{x-parcial-y}), the asymptotic
conditions (\ref{condi3}), and the
expressions (\ref{term(1)}), (\ref{term(6)}),
and eliminating terms with vanishing
contribution, we obtain,
\\
\begin{equation}
\int_{\partial\Sigma} dS_{a}^{(y)} Q^{a} =
\int_{\partial\Sigma} dS_{a}^{(y)} \,
h^{0ac}_{(y)} h^{0bd}_{(y)} \,
\sum_{e,f=1}^{3}
[ \, W^{e}_{c} W^{f}_{d}
\frac {\partial A_{ef}} {\partial y^{b}} -
     W^{e}_{d} W^{f}_{b}
\frac {\partial A_{ef}} {\partial y^{c}} \, ],
\label{term(7)}
\end{equation}
\\
and,
\\
\begin{equation}
\int_{\partial\Sigma}
\frac {1}{r^2}
\sum_{a=1}^{3} dS_{a}^{(y)} V^{a} =
\int_{\partial\Sigma}
\frac {1}{r^2}
\sum_{a,b,c,e,f=1}^{3} dS_{a}^{(y)}
A_{ef} \,
[ - y^{b} \delta^{ac}
W^{e}_{c} W^{f}_{b} +
\frac {(2\alpha - 1)} {r^2}
y^{a} y^{b} y^{c} W^{e}_{c} W^{f}_{b} ].
\nonumber \\
\label{term(13)}
\end{equation}
\\
We will need the following eqs.,
\\
\begin{eqnarray}
\quad \quad
\frac {\partial A_{ab}} {\partial y^{c}} =
\sum_{d=1}^{3}
( W^{d}_{c}
+ \frac {\partial \eta^{d}} {\partial y^{c}} )
\frac {\partial A_{ab}} {\partial x^{d}},
\quad \quad
dS_{a}^{(y)} = 
\sum_{b=1}^{3} W_{a}^{b} dS_{b}^{(x)},
\nonumber \\
\nonumber \\ 
\sum_{b=1}^{3}
W^{a}_{b} y^{b} = x^{a} - \eta^{a},
\quad \quad
h^{0ef}_{(x)} = 
W_{a}^{e}  W_{c}^{f} h^{0ac}_{(y)}.
\label{ecuaciones} 
\end{eqnarray}
\\
Introducing eqs. (\ref{ecuaciones}) in the
eqs. (\ref{term(7)}), (\ref{term(13)}),
and eliminating terms
with vanishing contribution, we find,
\\
\begin{eqnarray}
\int_{\partial\Sigma}
\sum_{a=1}^{3} dS_{a}^{(y)} Q^{a} & = &
\int_{\partial\Sigma} dS_{a}^{(y)} \,
h^{0ac}_{(y)} h^{0bd}_{(y)} \,
\sum_{e,f=1}^{3}
[ \, W^{e}_{c} W^{f}_{d}
\frac {\partial A_{ef}} {\partial y^{b}} -
     W^{e}_{d} W^{f}_{b}
\frac {\partial A_{ef}} {\partial y^{c}} \, ]
\nonumber \\
\nonumber \\
& = &
\int_{\partial\Sigma}
 dS_{a}^{(x)} h^{0ac}_{(x)} h^{0bd}_{(x)}
[ \, 
  \frac {\partial A_{cd}} {\partial x^{b}} -
  \frac {\partial A_{bd}} {\partial x^{c}} \, ],
\nonumber \\
\nonumber \\
\label{Q} 
\end{eqnarray}

\begin{eqnarray}
\int_{\partial\Sigma}
\sum_{a=1}^{3}
\frac {dS_{a}^{(y)}}{r^2} V_{a} & = &
\int_{\partial\Sigma}
\sum_{a,b,c,e,f=1}^{3}  
\frac {dS_{a}^{(y)}}{r^2} A_{ef} \,
[ - y^{b} \delta^{ac} W^{e}_{c} W^{f}_{b}
+ \frac {(2\alpha - 1)} {r^2} 
   y^{a} y^{b} y^{c} W^{e}_{c} W^{f}_{b} ]
\nonumber \\
\nonumber \\
& = & \int_{\partial\Sigma}
\sum_{a,b,c=1}^{3} 
\frac {dS_{a}^{(x)}} {r^2} \,
[ \, - x^{b} \delta^{ac} A_{bc}
+ \frac {(2\alpha - 1)} {r^2}
  x^{a} x^{b} x^{c} A_{bc} \, ].
\label{V} 
\end{eqnarray}
\\
Substituting (\ref{Q}) and (\ref{V})
in (\ref{masatotal(2)}), and
comparing with the eq. (\ref{masa1-asim}), 
we obtain,

\begin{eqnarray}
16 \pi (1-\alpha) M^{(2)}_{ADM\alpha} & = &
\int_{\partial\Sigma}
\sum_{a=1}^{3} dS_{a}^{(y)} \,
[ \, Q^{a} +
\frac {\alpha}{(1 - \alpha)^{2} r^2}
 V^{a} \, ]
\nonumber \\
\nonumber \\
& = & \int_{\partial\Sigma}
\sum_{a=1}^{3} dS_{a}^{(x)} \,
[ \, R_{(1)}^{a} + J_{(1)}^{a} \, ] =
16 \pi (1-\alpha) M^{(1)}_{ADM\alpha}.
\end{eqnarray}
\\
this proves the theorem.
\\
\\
Thus, we have obtained a natural and well
defined expression for the ADM mass for
the class of A.F.D.A.$\alpha$ spacetimes
which reproduces the intuitive result in
the case of the global monopole solution.
It might seem that the fact that these
solutions correspond to a negative value
of this mass \cite{Vilenkin2} is somehow
puzzling and rather different from the
A.F. case, where, under appropriate conditions
\cite{Schoen-Yau},\cite{witten}, the ADM
mass has been shown to be non-negative.
We can now understand this by noting that, 
both in the A.F.D.A.$\alpha$ case and the
A.F. case, the definition of the mass
comes about through the comparison of the
given 3- metric with a standard 3- metric
(which for the A.F. case, is taken to be the
standard flat metric, and for the A.F.D.A.
$\alpha$ case, is chosen to be $h^0_{ab}$).
The resulting mass is then a measure of the
degree to which the two metrics differ, and
 the issue of whether or not the mass is
non-negative can be restated as the
issue of whether or not one has chosen
the standard metric appropriately.
This choice could, then, be thought to be
a crucial one, however, we must note that
changing the standard metric (within a class)
would correspond to the addition of a constant
to our definition of the mass, a clearly
irrelevant change, to which we are, in fact,
accustomed in non-gravitational physics.
The relevant issue is, of course, whether or
not the mass is bounded from below,
and on this point we have very little to say,
 except that, in a specific theory,
like for example the global monopole sector
of the $O(3)$ model described in Sec. 1,
which corresponds to a specific value of 
$\alpha$, we expect that there will be a static solution, 
and that it will
 correspond to the minimum of $M_{ADM \alpha}$
within the phase space of regular
A.F.D.A. $\alpha$ initial data for that theory.
That this solution is an extremum of 
$M_{ADM \alpha}$, actually follows from the
same type of analysis as that carried out
in \cite{us1}, where it was shown that, for 
A.F. solutions of the EYM theory to be static,
it must correspond to extrema of $M_{ADM }$
( actually, in that case the solutions were 
extrema of $M_{ADM}$ at fixed Q). That the
static global monopole does actually
correspond to the global minima, would be
difficult to prove, but it is a natural
expectation, given that there
seems to be no candidate for a configuration
that would result from the decay of these
solutions. Needless to say, a real proof of
the lower boundness of $M_{ADM \alpha}$ for
theories for which the matter fields are
only required to satisfy certain energy
conditions would be very desirable.

\medskip
\section{Discussion}
\setcounter{equation}{0}
\medskip

We have introduced a class of spacetimes
(the A.F.D.A.$\alpha$ class)
characterized by an asymptotic behavior
compatible with the $1/r^2$ fall off rate
for the energy density, that is natural for
global monopoles, but could
include more general situations.
We have argued that this class allows
naturally the introduction of null
infinity in the form of $\J^{-}$ and $\J^{+}$,
but does not seem to allow the introduction
of spatial infinity $\imath^{0}$. We have,
nevertheless, been able to generalize the
definition of the ADM mass for this class
of spacetimes. It is natural, then, to expect
that, since the standard A.F.D.A. $\alpha$
spacetime, which we use to define the class,
possesses, not only a time translation
isometry, but also a full rotational
isometry group, one could also obtain
a well defined expression for the canonical
angular momentum $J$, whose formula would be
identical to the corresponding one for
A.F. spacetimes, but with the quantities
associated with the flat metric replaced
by the corresponding quantities associated
with $h^0_{ab}$. It is also gratifying that
one can introduce for these spacetimes
the notions of $\J^{-}$ and $\J^{+}$, which
allows, in turn, the definition of
black holes in A.F.D.A. $\alpha$ spacetimes.
The black hole region $B$ is given by
$B = M - J^{-}(\J^{+})$
(here $J^{-}$ indicates the causal past).
Moreover, the consideration of stationary
black holes, and their perturbations in the
same fashion as in \cite{us1} would 
straightforwardly yield the first law of black hole
thermodynamics in the A.F.D.A. $\alpha$ class,
\\
\begin{equation}
\delta M + V\delta Q -\Omega \delta J =
\frac {1}{8\pi (1-\alpha)} \kappa
\delta A \label{bh}
\end{equation}
\\
where $\Omega$ is the angular velocity of
the horizon, $\kappa$ is the surface gravity
of the horizon (these quantities being defined
in the same way as in the A.F. case.),
and A is the area of the black hole horizon.
The variations $\delta$ refer to perturbations
within the corresponding phase space, about
stationary black hole
(See \cite{us1} for details).
\\

Finally, we would like to point out the fact that,
in analogy with what happens in the A.F.
case, where the asymptotic symmetry group
is larger than the symmetry group of
the standard spacetime ( i.e., includes the
supertranslations not present in Minkowski
spacetime), in the present case, one also
expects the asymptotic symmetry
group of the A.F.D.A. $\alpha$ class
of spacetimes to be larger than that of
the standard A.F.D.A. $\alpha$
spacetime, which consists only of rotations
and time translations. Looking at the
general form of the coordinate
transformation that preserves the asymptotic
behavior of the A.F.D.A. $\alpha$ metric,
one is led to expect the asymptotic
symmetry group to include, among other
transformations, translations not present
 in the standard spacetime. It seems, however,
that substantial progress will be required
before one can tackle this issue rigorously,
due to the impediments that the program faces
given the fact that we are not even
able to define the point at spatial
infinity $\imath^{0}$.

\medskip
\section{Acknowledgments}
\setcounter{equation}{0}
\medskip

We wish to thank Prof. P. Chru\'{s}ciel, Prof. R. M. Wald
and Prof. J. Guven
for helpful discussions, and to Prof. R. Sorkin 
for useful comments
and for carefully reading the manuscript.
D.S. would like to acknowledge partial support
from DGAPA-UNAM project IN105496.
\\

\eject
\end{document}